\documentclass[12pt,epsf]{article}

\usepackage{amsmath}
\usepackage{amsthm, amssymb}
\usepackage{amsfonts}
\usepackage{fixltx2e}%
\usepackage{booktabs} 
\usepackage{url}
\usepackage{graphicx}

\usepackage{xcolor}
\usepackage{framed}
\definecolor{shadecolor}{gray}{0.95}

\usepackage{mathrsfs}
\usepackage{setspace}
\usepackage{subfigure}
\usepackage{slashed}
\usepackage{mathtools}
\usepackage{amssymb}
\usepackage{amsfonts}
\usepackage{tensind}
\usepackage{float}
\usepackage{microtype}%
\usepackage{amsmath}%
\usepackage{setspace}
\usepackage{amsthm}%
\mathtoolsset{showonlyrefs}
\usepackage[utf8]{inputenc}%
\usepackage{csquotes}
\usepackage{textcomp}%
\usepackage[english]{babel}%
\usepackage{empheq}

\usepackage{bm}

\newenvironment{definition}[1][Definition:]{\begin{trivlist}
\item[\hskip \labelsep {\bfseries #1}]}{\end{trivlist}}

\usepackage{tikz}
\usetikzlibrary{decorations.markings}
\usetikzlibrary{positioning,arrows,matrix,patterns}
\usetikzlibrary{calc}
\usetikzlibrary{decorations.pathmorphing}
\usepackage{tikz-cd}

\renewcommand{\bar}[1]{\overline{#1}} 
\renewcommand{\hat}[1]{\widehat{#1}} 

\renewcommand{\vec}[1]{\mathbf{#1}}

\newcommand{\D}{\,\mathrm{d}}

\DeclareMathAlphabet{\mathpzc}{OT1}{pzc}{m}{it}

\newcommand{\nbox}{{\,\lower0.9pt\vbox{\hrule \hbox{\vrule height 0.2 cm \hskip 0.19 cm \vrule height 0.2 cm}\hrule}\,}}

\def\href#1#2{#2}

\usepackage{color}

\usepackage{xcolor}
\usepackage[bookmarks,colorlinks,urlcolor=blue,colorlinks = true,
linkbordercolor = {white},
citebordercolor = {white},
citecolor = {blue},
linkcolor = {blue}]{hyperref} 

\begin{document}
\begin{titlepage}
	\begin{flushright}
		
		\texttt{ MPP-2019-179 }
		
	\end{flushright}

\begin{NoHyper}
\hfill
\vbox{
    \halign{#\hfil         \cr
           } 
      }  
\vspace*{20mm}
\begin{center}
{\Large \bf Quasi-local conserved charges in General Relativity}

\vspace*{15mm}
\vspace*{1mm}
Henk Bart 
\vspace*{1cm}
\let\thefootnote\relax

{
	\textit{Max-Planck-Institut f\"ur Physik,\\
	F\"ohringer Ring 6, 
	80805 M\"unchen, Germany\\}
	\quad \\
	\href{mailto:hgjbart@mpp.mpg.de}{hgjbart@mpp.mpg.de}
}

\vspace*{1cm}
\end{center}

\begin{abstract} 
A general prescription for constructing quasi-local conserved quantities in General Relativity is proposed. The construction is applied to BMS symmetry generators in Newman-Unti gauge, so as to define quasi-local BMS charges. It is argued that the zero mode of this BMS charge is a promising definition of quasi-local energy.  
\end{abstract}
\end{NoHyper}

\end{titlepage}
\vskip 1cm
\begin{spacing}{1.15}

	\tableofcontents

\newpage

\section{Introduction and summary of results}
Consider a closed spacelike two-surface $B$ in a four dimensional spacetime $M$. Then a question in General Relativity is:
\begin{center}
	What is a sensible notion of energy in the region enclosed by $B$?
\end{center}
Such a notion of energy will be referred to as a \emph{quasi-local energy}. 

The history of defining quasi-local energy started with the observation that a local notion of energy and momentum -- a stress energy tensor -- does not exist for the gravitational field in General Relativity. This follows directly from the equivalence principle and is therefore a general property of  diffeomorphism covariant theories. 

However, quasi-local notions of conserved quantities are not ruled out by the equivalence principle. They are expected to be useful for various reasons\footnote{See \cite{Szabados:2009eka} for an overview.}. For example, they could provide a more detailed characterisation of states of the gravitational field than the globally defined quantities. Furthermore, they are important from the point of view of applications, such as formulating and proving various conjectures\footnote{One example is the \emph{Hoop conjecture}, which is a  criterion for when a black hole forms under gravitational collapse. In order to formulate this conjecture more precisely, a good notion of quasi-local energy is needed.} in General Relativity, as well as formulating the laws of black hole thermodynamics \cite{Ashtekar:2004cn}.

Therefore, the hope has been that it will be possible to construct a quasi-local energy. There is some justification for this hope, because a variety of such quantities have appeared in the literature. Examples are the Komar mass \cite{Komar:1958wp},  Misner-Sharp energy \cite{Misner:1964je},  Hawking energy \cite{Hawking:1968qt}, Bartnik mass \cite{Bartnik:1989zz}, Brown-York energy \cite{Brown:1992br} and the Wang-Yau mass \cite{Wang:2008jy} among many others. See \cite{Szabados:2009eka} for an overview.

A problem is that the applicability of the known quasi-local quantities breaks down at one point or another. This happens, for instance, because the quantity is only defined in special cases, or because the quantity is (physically) ill-behaved outside of a class of solutions.  To indicate the severity of the problem, let us point out that the most well-known notion of quasi-local energy by Brown and York \cite{Brown:1992br} does not in general vanish in the Minkowski spacetime.

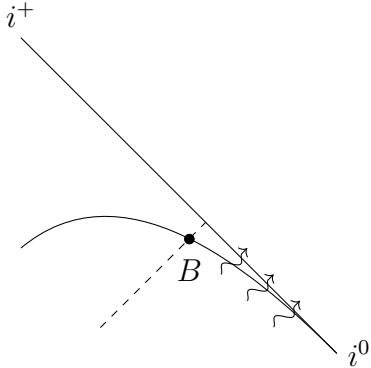
\begin{figure}[h!]
	\centering
	\begin{tikzpicture}[ scale=0.7]
	;
	\coordinate (i+) at (0,6) {};
	\coordinate (i0) at (6,0) {};
	\node (shift) at (0.5,0.5) {};
	
	\draw (i0) -- node [at start,  right] {$i^0$} node [at end,  above]{$i^+$} (i+);
	\draw (0,2) to[out=40,in=135] (i0);

	\draw[dashed] ($(2,0)+0.49*(-1,1)$) -- ($(4,2)+0.49*(-1,1)$);
\draw[black,fill=black] (3.2,2.172) circle (.5ex) node [below ] {$\overset{ }{B}$ };
\draw[->,decorate,decoration=snake]  ($(5.3,0)+1*(-0.5,0.5)$) to node[at end,right] { } ($(5.3,0)+(shift)+1*(-0.5,0.5)$) ;
\draw[->,decorate,decoration=snake]  ($(5.3,0)+2*(-0.5,0.5)$) to node[at end,right] { } ($(5.3,0)+(shift)+2*(-0.5,0.5)$) ;
\draw[->,decorate,decoration=snake]  ($(5.3,0)+3*(-0.5,0.5)$) to node[at end,right] { } ($(5.3,0)+(shift)+3*(-0.5,0.5)$) ;
	\end{tikzpicture}
	\caption{The Bondi mass at null infinity may be thought of as a quasi-local charge in the sense that it is given by the total energy of the spacetime minus the energy of the radiation that went out at earlier times. Here we think of the quasi-local region as the region enclosed by a sphere $B$ in a spacelike slice that approaches null infinity.}
	\label{fig:WZ-infty}
\end{figure}

The goal of the present paper is to provide a general framework for constructing quasi-local conserved quantities, and a notion of quasi-local energy that does not suffer from some of the problems referred to above. Our starting point will be the construction of conserved quantities associated with asymptotic symmetries at null infinity by Wald and Zoupas \cite{Wald:1999wa}. Though at null infinity, these charges may be thought of as quasi-local charges if one thinks of a cut at null infinity as a sphere ($B$ in \autoref{fig:WZ-infty}) in a spacelike slice that is sent outwards with the speed of light. The quasi-local region then contains the \emph{Bondi energy}: the total energy of the spacetime minus the energy of the radiation that was sent out at earlier times. 

The construction of Wald and Zoupas provides a notion of ``conserved quantity'' in situations where a Hamiltonian associated with a symmetry generator does not exist. This is the case at null infinity, because unlike at spacelike infinity, the quantity that would be the Hamiltonian is not conserved due to in- or outgoing radiation. The situation is similar in the bulk of a spacetime; the would-be Hamiltonian is also not conserved due to the presence of radiation or matter. However, there it appears that the construction of Wald and Zoupas cannot be applied. This is because the defining conditions of the Wald-Zoupas charges are tailored to the special case of null infinity. 

Nevertheless, given the quasi-local nature of the Wald-Zoupas charges, the hope has been that their construction provides clues about how to define conserved quantities in the bulk of the spacetime. We shall argue that, indeed, a modification of their procedure leads to a well-defined notion of quasi-local conserved charges. The first purpose of the present paper is thus to modify the construction of Wald and Zoupas so that the ``conserved charges'' exist more generally, and in particular in the bulk of a spacetime. This then provides a new definition of \emph{quasi-local conserved charges} associated with generators of diffeomorphisms. 

Let us outline the technical steps that we will take in terms of the construction of Wald and Zoupas.
\begin{enumerate}
	\item[] The construction of Wald and Zoupas is essentially a proposal for a correction term $\bm{\Theta}(\phi,\delta \phi)$, which is added to the defining equation of the Hamiltonian to guarantee the existence of a solution. Here $\phi$ denote the fields of the theory and $\delta \phi$ denote variations thereof. 
	
	One of the defining conditions of the correction term  $\bm{\Theta}(\phi,\delta \phi)$  at null infinity is that it vanishes for every \emph{stationary} solution $\phi$. This condition makes sense at null infinity, because the quantity $\bm{\Theta}(\phi,\mathcal{L}_\xi \phi)$ is equal to the flux of the charge associated with $\xi$, and at null infinity of stationary spacetimes there is no radiation. In the bulk of a stationary spacetime, however, there may exist other types of matter which in general account for the non-vanishing of the flux. Therefore, the \emph{stationarity condition} is not applicable in the bulk. We propose instead the following defining condition of $\bm{\Theta}(\phi,\delta \phi)$.

	Consider an auxiliary hypersurface ${}^{(3)}\!B$. We define $\bm{\Theta}(\phi,\delta \phi)$ on ${}^{(3)}\!B$ by the condition that $\bm{\Theta}(\phi,\delta \phi)$, for variations $\delta$ that respect a given type of boundary conditions $X$ on ${}^{(3)}\!B$, integrates to zero on every closed spacelike two-surface $B$ contained in ${}^{(3)}\!B$. For vector fields tangent to ${}^{(3)}\!B$, this defines an associated quasi-local conserved charge with respect to boundary conditions $X$.
	
	This condition is, however, not sufficient, since it defines $\bm{\Theta}(\phi,\delta \phi)$ up to a term which is invariant under variations that preserve the boundary conditions $X$.
	Moreover, the resulting charge at a closed spacelike two-surface $B$ may be ill-defined, because its definition depends on the choice of auxiliary hypersurface ${}^{(3)}\!B$. We refer to the freedom in the choice of $\bm{\Theta}(\phi,\delta \phi)$ as a choice of \emph{reference term}.
	
	To make our proposal well-defined, we shall impose consistency conditions on the reference term, so that the resulting charge can be interpreted unambigiously as a quasi-local charge at $B$ (independent of the choice of auxiliary ${}^{(3)}\!B$).
	We further restrict the freedom in the choice of reference term by introducing \emph{orthogonality} and \emph{zero point conditions}.
	
	A  correction term $\bm{\Theta}(\phi,\delta \phi)$ that satisfies the conditions mentioned above defines -- through the usual procedure \cite{Wald:1999wa} -- a quasi-local conserved charge.
\end{enumerate}

The second purpose of the present paper is to apply our construction to BMS symmetries \cite{Bondi:1962px,Sachs:1962wk,Sachs:1962zza}, so as to define \emph{quasi-local BMS charges}. In the literature,  attempts at defining quasi-local BMS charges have been made. See\footnote{See also \cite{Flanagan:2015pxa,Donnay:2015abr,Compere:2016jwb,Compere:2018ylh,Avery:2016zce} for computations of BMS (type) charges in the linearised theory, and \cite{Hawking:2016msc,Hawking:2016sgy,Haco:2019ggi,Averin:2016ybl,Bousso:2017rsx,Gomez:2017ioy,Donnay:2018ckb} for discussions about and against their relevance in questions concerning black hole entropy.} e.g. \cite{Hawking:2016msc,Hawking:2016sgy,Haco:2018ske,Chandrasekaran:2018aop,Haco:2019ggi,Godazgar:2018qpq,Godazgar:2018dvh,Godazgar:2019dkh}. However, a drawback of these constructions is that they are not derived from a general framework such as the one  developed in this work. In addition, several ambiguities are left untreated, such as the definition of BMS generators in the bulk of a spacetime, which we now comment on.

BMS symmetries are asymptotic symmetries of asymptotically flat spacetimes. They are not a priori defined in the bulk of a spacetime. Therefore, in order to be able to define quasi-local BMS charges, one has to provide a method for extending such symmetries into the bulk. Bulk extensions of BMS generators  exist in the literature, such as the extensions in Bondi gauge \cite{Bondi:1962px} and Newman-Unti gauge \cite{Newman:1962cia}. These are uniquely determined by the requirement that BMS generators in the bulk preserve the given gauge conditions. However, a problem with extensions of this kind is that the  gauge choice is essentially arbitrary, and that the generators depend on this gauge. This makes it non-trivial to construct gauge invariant charges. 

We do not solve the issue of gauge dependence of the BMS charge. However, we  provide in a separate paper \cite{bart2019} a justification for why the BMS generators in Newman-Unti gauge are physically preferred. Namely, that BMS generators in Newman-Unti gauge are connected to the \emph{gravitational memory effect}.  

We show that in Newman-Unti gauge our quasi-local BMS charges have the following properties.

\begin{enumerate}
	\item[(i)] The charges vanish in the Minkowski spacetime.
	\item[(ii)] The charges coincide asymptotically at null infinity with the BMS charges constructed by Wald and Zoupas.
	\item[(iii)] At the outer horizon of a Reissner-N\"ordstrom black hole, the zero mode ($f=1$) of the gravitational part of the BMS charge is the \emph{irreducible mass} of the black hole.
\end{enumerate}
Since these are pragmatic criteria that a useful definition of quasi-local energy is expected to satisfy (see e.g. \cite{Szabados:2009eka} for a list of criteria), we shall put the zero mode BMS charge forward as a new definition of quasi-local energy.

\subsubsection*{Organisation}
In \autoref{sec:reduced-phase-space} we review the construction of a Hamiltonian on spacetimes with a boundary. This serves as a preparation for \autoref{sec:conserved-charges}, where a general prescription for defining quasi-local charges (on a hypersurface ${}^{(3)}\!B$ which is not necessarily a boundary of the spacetime) in diffeomorphism covariant theories is provided. In \autoref{eq:quasi-local-GR}, we construct quasi-local charges in General Relativity. In \autoref{sec:quasi-local-BMS}, we consider BMS generators in Newman-Unti gauge, we evaluate the corresponding charges, and we discuss how the zero mode BMS charge may serve as a definition of quasi-local energy. Possible directions for future work are discussed in \autoref{sec:discussions}.

\section{Quasi-local charges on a boundary}\label{sec:reduced-phase-space}
Consider a manifold $M$ with boundary $\partial M$. Let $B$ be a closed spacelike codimension two-surface in $\partial M$. Then a quasi-local conserved quantity on $B$ may be defined as a Hamiltonian associated with a vector field that is tangent to $\partial M$ \cite{Brown:1992br,Iyer:1995kg}. The construction of a Hamiltonian on the boundary of a spacetime forms a basis for the ideas presented in this paper. Therefore, we begin with a review of this construction. We follow \cite{Iyer:1995kg} and we adapt to the notation that boldface symbols are differential forms on the spacetime.

On an $n$-dimensional manifold $M$ with boundary $\partial M$, we consider a diffeomorphism covariant theory defined by the action
\begin{equation}\label{eq:general-action}
S_{X} = \int_M \mathbf{L} - \int_{\partial M} \mathbf{B}_X.
\end{equation}
Here $\mathbf{L}$ is a Lagrangian $n$-form and $\mathbf{B}_X$ is an $(n-1)$-form on $\partial M$ associated with boundary conditions $X$. For variations $\delta$ of the fields $\phi$ that respect the boundary conditions $X$, the variation of the action,
\begin{equation}\label{eq:variation-lagrangian}
\delta S_X = \int_{M} \big[\mathbf{E}(\phi)\delta \phi + \D \bm{\theta}(\phi,\delta \phi)\big]-\int_{\partial M} \delta \bm{B}_X,
\end{equation}
yields the equations of motion $\mathbf{E}=0$ when the boundary term satisfies
\begin{equation}\label{eq:boundary-term-variation}
\delta \bm{B}_X(\phi) = \bar{ \bm{\theta}}(\phi,\delta \phi)|_{\partial M} - \bar{\D \bm{\mu}}(\phi,\delta \phi)|_{\partial M}.
\end{equation}
Here $\bm{\mu}$ is a $2$-form and $\bar{\bm{\theta}}(\phi,\delta \phi)|_{\partial M}$ denotes the pull-back of  $\bm{\theta}(\phi,\delta \phi)$ onto the boundary. 

An example of such a theory is General Relativity on a manifold with a timelike boundary $\partial M$ equipped with \emph{canonical boundary conditions} $X$ defined as follows.

\begin{definition}[Canonical boundary conditions:]\label{def:canonical-boundary-conditions}
Boundary conditions $X$ are called \emph{canonical} if the fields whose variation appears in $\bm{\theta}(\phi,\delta \phi)$ are held fixed on $\partial M$. 
\end{definition}
\noindent
An action for this theory is the Einstein-Hilbert action supplemented with the Gibbons-Hawking-York boundary term  \cite{York:1972sj,Gibbons:1976ue,York:1986lje}, given by
\begin{equation}\label{eq:canonical-action}
S = \frac{1}{16\pi}\int_M \bm{R}  + \frac{1}{8\pi} \int_{\partial M} \big( \bm{K} - \bm{K}_0\big) .
\end{equation}
Here $\bm{R}$ denotes the Ricci scalar and $\bm{K}$ is the trace of the extrinsic curvature density of the (timelike) boundary $\partial M$. The boundary three-form $\bm{K}_0$ is any functional of the boundary metric. It represents an ambiguity of the action for this choice of boundary conditions. The freedom in choosing $\bm{K}_0$ may be viewed at as a choice of zero point for the Hamiltonian, to which we turn our attention now.

We review the construction of a Hamiltonian on $M$ in the covariant phase space formalism \cite{Iyer:1995kg,Harlow:2019yfa}. For a theory of the form \eqref{eq:variation-lagrangian}, the symplectic two-form density is given by the variational exterior derivative of the canonical one form $\bm{\theta}(\phi,\delta \phi)$. Given two independent field variations $\delta_1 \phi$ and $\delta_2 \phi$, that is,
\begin{equation}\label{eq:symplectic-2-form}
\bm{\omega}(\phi,\delta_1 \phi, \delta_2 \phi) := \delta_1 \bm{\theta}(\phi,\delta_2 \phi) - \delta_2 \bm{\theta}(\phi,\delta_1 \phi).
\end{equation}
Consider a foliation of $M$ given by achronal slices $\Sigma_t$ (labeled by a parameter $t$), which intersect $\partial M$ orthogonally in compact spacelike $(n-2)$-dimensional surfaces $C_t$. Then the (pre-)symplectic two form is given by
\begin{equation}\label{eq:Omega_Sigma}
\Omega_{\Sigma_t}(\phi,\delta_1 \phi, \delta_2 \phi) := \int_{\Sigma_t}\bm{\omega}(\phi,\delta_1 \phi, \delta_2 \phi).
\end{equation}

Let $\xi$ be any vector field on $M$ that is tangent to $\partial M$. Then we say that a real-valued function $H_\xi$ on the covariant phase space is a \emph{Hamiltonian conjugate to $\xi$} if for all variations of the field that respect the boundary conditions $X$,
\begin{equation}\label{eq:def-hamiltonian}
\delta H[\xi] = \Omega_{\Sigma_t}(\phi,\delta \phi,\mathcal{L}_\xi \phi).
\end{equation}
Here $\mathcal{L}_\xi$ denotes the Lie-derivative with respect to $\xi$. As shown in \cite{Iyer:1995kg}, for variations $\delta \phi$ that satisfy the linearized equations of motion, and for on-shell solutions $\phi$, it holds true that
\begin{equation}
\Omega_{\Sigma_t}(\phi,\delta \phi,\mathcal{L}_\xi \phi) =  \int_{C_t}\delta \vec{Q}[\xi] - \xi \cdot \bm{\theta}(\phi,\delta \phi),
\end{equation}
where $\bm{Q}[\xi]$ is the \emph{Noether charge two-form}. Then, using \eqref{eq:boundary-term-variation} and the \emph{assumption} that the pull-back $\bar{\bm{\mu}}|_{C_t}$ of $\bm{\mu}$ to $C_t$ vanishes, a solution to \eqref{eq:def-hamiltonian} exists and is given by
\begin{equation}\label{eq:Hamiltonian}
H_X[\xi] = \int_{C_t}\vec{Q}[\xi] - \xi \cdot \mathbf{B}_X.
\end{equation}
This is a quasi-local conserved quantity defined on the boundary $\partial M$. 

Notice that $\bm{B}_X$ is in general determined up to a three-form which depends only on the boundary data $X$. We shall return to this freedom of choosing a \emph{reference term} ($\bm{K}_0$ in \eqref{eq:canonical-action}) later.

The following can be said about \eqref{eq:Hamiltonian}. General Relativity satisfies the requirements for the existence of $H_X[\xi]$ for canonical boundary conditions $X$. When the gravitational reference term $\bm{K}_0$ in \eqref{eq:canonical-action} is the Hawking-Horowitz-Hunter reference term \cite{Hawking:1996ww,Hawking:1995fd}, the Hamiltonian \eqref{eq:Hamiltonian} conjugate to unit time translations at spacelike infinity is the ADM mass \cite{Arnowitt:1959ah} plus possibly additional contributions from long range matter fields \cite{Wald:1993nt}. When $\partial \Sigma_t$ is an inner-boundary in the bulk of the spacetime, and $\xi$ is a unit time translation, \eqref{eq:Hamiltonian} is a generalisation of the Brown-York quasi-local energy \cite{Brown:1992br}.

\section{General definition of quasi-local charges}\label{sec:conserved-charges}
The function $H_X[\xi]$ constructed in \eqref{eq:Hamiltonian} is a true Hamiltonian function on the phase space, only if the phase space incorporates the boundary conditions $X$. There are, however, situations where it is desired to consider a more general class of solutions that violate the boundary conditions $X$, but where a quantity like $H_X[\xi]$ is still physically meaningful.

One example is null infinity as a boundary of asymptotically flat spacetimes. On the phase space consisting of all asymptotically flat spacetimes, a solution to \eqref{eq:def-hamiltonian} does not exist at null infinity. However, the \emph{Bondi mass} that exists as a Hamiltonian function on the reduced phase space where in- and outgoing radiation is excluded, turns out to be physically relevant on the original phase space too \cite{Wald:1999wa}. Only, it is not conserved when radiation enters or leaves through null infinity. This observation indicates that it could be useful have a procedure for constructing ``conserved quantities'', even though strictly speaking the quantities are not Hamiltonian functions on the phase space.

The goal of this section is to provide a prescription for constructing a ``conserved quantity'' assocoiated with a vector field $\xi$ on an arbitrary closed spacelike two surface $B$ in $M$. Our prescription is based on  the framework of Wald and Zoupas for constructing ``conserved quantities'' in diffeomorphism covariant theories in situations where a Hamiltonian does not exist. Their prescription leads to a well-defined notion of conserved charges at null infinity. However, it is not in general applicable in the bulk of a spacetime.  We shall provide a modification of their prescription that is applicable in a more general context and in particular in the bulk of a spacetime.

The organisation of this section is as follows. In \autoref{sec:review-WZ}, we review the construction of Wald and Zoupas and we explain why it is not applicable in the bulk of a spacetime. In \autoref{sec:general-definition}, we propose a modification of their construction that is applicable more generally. In \autoref{sec:consistency} and \autoref{sec:orthogonality-zero-point} we impose consistency conditions. A summary of our proposal is provided in  \autoref{sec:summary-proposal}.

\subsection{The Wald-Zoupas correction term}\label{sec:review-WZ}
Consider a hypersurface ${}^{(3)}\!B$ in $M$. (Our notation is adapted to the situation where the spacetime dimension is $n=3+1$.) Let $\bm{\Theta}$ be a symplectic potential for the pull back $\bar{\bm{\omega}}$ of $\bm{\omega}$ onto ${}^{(3)}\!B$. That is, $\bm{\Theta}$ satisfies
\begin{equation}\label{eq:pull-back-omega}
\bar{\bm{\omega}}(\phi,\delta_1 \phi, \delta_2 \phi) = \delta_1 \bm{\Theta}(\phi,\delta_2 \phi) - \delta_2 \bm{\Theta}(\phi,\delta_1 \phi).
\end{equation}
Wald and Zoupas \cite{Wald:1999wa} then define a ``conserved quantity'' $\mathcal{H}[\xi]$ conjugate to a vector field $\xi$ tangent to ${}^{(3)}\!B$ as a solution to the equation 
\begin{equation}\label{eq:def-conserved-quantity}
\delta \mathcal{H}[\xi] =  \Omega_{\Sigma}(\phi,\delta \phi,\mathcal{L}_\xi \phi) + \int_{B} \xi \cdot \bm{\Theta}.
\end{equation}
Here $\Omega_{\Sigma}$ is defined by \eqref{eq:Omega_Sigma} in which $\Sigma$ is an achronal slice with a boundary at $B\subset {}^{(3)}\!B$. Thus, the idea of Wald and Zoupas is to introduce $\bm{\Theta}$ as a \emph{correction term} in the defining equation of the Hamiltonian \eqref{eq:def-hamiltonian}, such that a solution exists, even in situations where originally it does not.

Note that $\bm{\Theta}$ must be of the form
\begin{equation}\label{eq:form-of-Theta}
\bm{\Theta}(\phi,\delta \phi) = \bar{\bm{\theta}}(\phi,\delta \phi) - \delta \bm{W}(\phi),
\end{equation}
where $\bar{\bm{\theta}}$ is the pull-back of $\bm{\theta}$ onto ${}^{(3)}\!B$ and $\bm{W}$ is an arbitary three-form on ${}^{(3)}\!B$. It follows that a solution to \eqref{eq:def-conserved-quantity} is given by
\begin{equation}
\mathcal{H}[\xi] = \int_{B} \bm{Q}[\xi] - \xi\cdot \bm{W}.
\end{equation}
However, since $\bm{W}$ is essentially arbitrary, the above prescription is not well-defined. One must impose by hand a sensible condition or procedure to specify $\bm{W}$.

 In order to fix the ambiguity in $\bm{W}$, Wald and Zoupas \cite{Wald:1999wa} imposed the condition that 
 \begin{equation}\label{eq:WZ-condition}
 \bm{\Theta}(\phi,\delta \phi)=0,
 \end{equation}
 for every stationary spacetime $\phi$ and on-shell perturbation $\delta \phi$. They showed that in the limit where ${}^{(3)}\!B$ approaches null infinity, this condition uniquely defines $\bm{\Theta}$, and that it gives rise to the Bondi mass as the conserved charge associated with unit time translations.

The motivation to fix $\bm{\Theta}$ by the requirement that it vanishes on  stationary spacetimes is that $\bm{F}_\xi:=\bm{\Theta}(\phi,\mathcal{L}_\xi \phi)$ is the flux of the charge conjugate to $\xi$. I.e., for a submanifold $\Delta \subset {}^{(3)}\!B$, 
\begin{equation}
 \mathcal{H}[\xi]\big|_{\partial \Delta} = \int_{\Delta}\bm{F}_\xi.
\end{equation}
This means that the requirement \eqref{eq:WZ-condition} is physically justified at null infinity, because there is no in- or outgoing radiation (Bondi news) in stationary spacetimes.

\begin{figure}[h!]
	\centering
	\begin{tikzpicture}[ scale=0.7]	
		\draw[draw=white,fill=lightgray,opacity=0.5] (0,0) rectangle ++(1,8);
		
		\draw (-3,0) to[in=-135,out =60] (-1,3);
		\draw[line width = 3] (-1,3)  to[in=-135,out =45](2,5); 
		\draw (2,5) to[in=-90,out =45] node[at end,right] {${}^{(3)}\!B$} (3,8);
		
		\node (Delta) at (-2,5) {$\Delta $}; 
		\draw[->,thick,>=stealth] ($(Delta)+0.4*(1,-1)$) to[in=100,out =-45] ($(-0.2,3)+0.5*(-1,1)$);
		
		
		\node (F) at (7,5) {$\displaystyle\int_{\Delta} \bm{F}_{\xi} \neq 0$}; 
		
		\draw[->,thick,>=stealth] (3,0) -- node[at end,below] {$r$} (5,0);
		\draw[->,thick,>=stealth] (3,0) -- node[at end,left] {$t$} (3,2);
		
		\node[below] at (0,0) {$r_0$};
		\node[below] at (1,0) {$r_1$};
	\end{tikzpicture}
	\caption{A spacetime containing a stationary shell of matter between $r_0 < r < r_1$ (the grey rectangle). Each point in the figure represents a two-sphere at radius $r$ and time $t$. When the hypersurface ${}^{(3)}\!B$ intersects a region containing the matter, one in general expects that the flux $\bm{F}_\xi$ of the charge associated with a vector $\xi$ tangent to ${}^{(3)}\!B$ is non-vanishing.}
		\label{fig:non-zero} 
\end{figure}
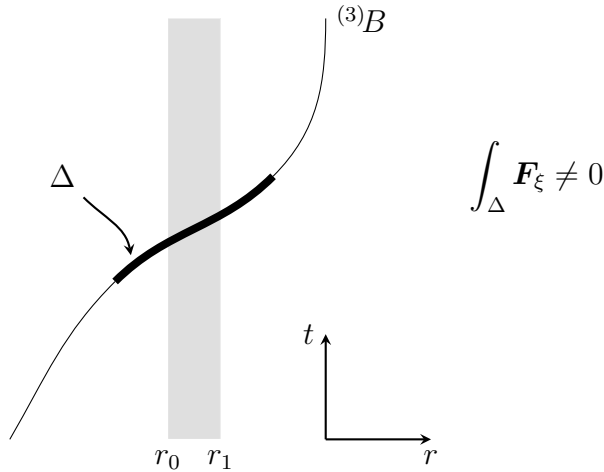

However, the requirement \eqref{eq:WZ-condition} is not physically justified\footnote{In \cite{Flanagan:2015pxa} the stationarity condition \eqref{eq:WZ-condition} is replaced by the requirement that $\Theta(\phi,\delta \phi)$ vanishes on null surfaces with vanishing shear and vanishing expansion.} when ${}^{(3)}\!B$ is a hypersurface in the bulk of a spacetime. Namely, even if $\phi$ is stationary, when ${}^{(3)}\!B$ intersects a (stationary) source of matter, one expects that the flux through ${}^{(3)}\!B$ is non-zero. Such a situation is depicted in \autoref{fig:non-zero}. We shall therefore propose an alternative method to specify $\bm{\Theta}$, which is also applicable in the bulk of a spacetime.

\subsection{Correction terms in the bulk}\label{sec:general-definition}
Here we propose a correction term $\bm{\Theta}(\phi,\delta \phi)$ in the bulk of a spacetime. The idea is the following. Instead of requiring that $\bm{\Theta}(\phi,\delta \phi)$ vanishes on a given class of spacetimes $\phi$, we shall require that it integrates to zero on $B$ for a type of variations $\delta \phi$.  How do we define a ``type of variation''? We consider boundary conditions $X$ on a ${}^{(3)}\!B$ that contains $B$. A variation of the type $X$ is then defined as a variation of the fields $\delta \phi$ that preserves the boundary conditions $X$. 

The resulting ``conserved quantity'' is by construction -- if it exists -- identical to the Hamiltonian \eqref{eq:Hamiltonian} on the reduced phase space that incorporates the boundary conditions $X$.  The difference with the previous section is that the boundary conditions do not constrain the phase space. They serve only to define the type of variations (processes) for which the flux through ${}^{(3)}\!B$ vanishes.

\begin{definition}\label{def:WZ-X}
	Let ${}^{(3)}\!B$ be a hypersurface in $M$. Choose $\bm{\Theta}(\phi,\delta \phi)$ in \eqref{eq:form-of-Theta} such that there exists an $(n-2)$-form $\bm{\mu}$, such that for every variation $\delta \phi$ that respects a given choice of boundary conditions $X$ on ${}^{(3)}\!B$,
	\begin{equation}\label{def:Theta}
	\bm{\Theta}(\phi,\delta \phi)= \bar{\D \bm{\mu}}(\phi,\delta \phi).
	\end{equation}
	Here $\overline{\D \bm{\mu}}$ denotes the pullback of $\D \bm{\mu}$ to ${}^{(3)}\!B$. Then at $B\subset {}^{(3)}\!B$, we shall call a solution $\mathcal{H}_X[\xi]$ to \eqref{eq:def-conserved-quantity} a \emph{quasi-local conserved charge with respect to boundary conditions X}.
\end{definition}
Different choices of boundary conditions $X$ yield different ``conserved charges'', each of which has its own physical interpretation. The role of the boundary conditions is to determine what part of the total charge (e.g. energy) is available to an outside observer which respects the boundary conditions $X$. The situation is similar in statistical thermodynamics, where different ensembles have different free energies. Thus, with the proposed definition, the problem of constructing quasi-local conserved quantities reduces to finding meaningful boundary conditions.

With the proposed choice of $\bm{\Theta}(\phi,\delta\phi)$, the ``conserved charge'' takes the form
\begin{equation}\label{eq:charge-general-0}
	\mathcal{H}_X[\xi] = \int_{B} \bm{Q}[\xi] -\xi \cdot \bm{B}_X,
\end{equation}
where $\bm{B}_X$ satisfies 
\begin{equation}\label{eq:theta-B}
\bar{\bm{\theta}}(\phi,\delta \phi)-\delta \bm{B}_X (\phi)= \bar{\D \bm{\mu}},
\end{equation} 
for variations $\delta$ that respect the boundary conditions $X$ on ${}^{(3)}\!B$. 

Notice, however, that \eqref{eq:charge-general-0} is not yet well-defined. Namely, $\bm{B}_X$ is defined up to the addition of a three form $\bm{B}_X^0$ such that 
\begin{equation}
\delta \bm{B}^0_X=0,
\end{equation}
 for variations $\delta$ that respect the boundary conditions $X$. We  refer to $\bm{B}_X^0$ as a \emph{reference term}, which we discuss momentarily.
 
In the remainder of this paper, unless stated otherwise, we choose $X$ to be \emph{canonical boundary conditions} as defined in \autoref{sec:reduced-phase-space}. We refer to the corresponding charge as the \emph{canonical quasi-local conserved charge}.

\subsection{Consistency of the reference term}\label{sec:consistency}
The correction term $\bm{\Theta}(\phi,\delta \phi)$ in \eqref{def:Theta} is  defined up to a choice of  \emph{reference term} $\bm{B}^0_X$ that satisfies $\delta \bm{B}^0_X=0$ for variations $\delta$ that respect the boundary conditions $X$. One has to remove this freedom by hand. 

In the previous section, in e.g. \eqref{eq:Hamiltonian}, there was a similar type of freedom. However, in contrast to the previous section, there are now consistency conditions that restrict the freedom in the choice of reference term $\bm{B}^0_X$. 

\subsubsection*{Tangent condition}
Namely, in contrast to \eqref{eq:Hamiltonian}, the boundary term  $\bm{B}_X$  in \eqref{eq:charge-general-0} is defined simultaneously on any auxiliary hypersurface ${}^{(3)}\!B \supset B$ that is tangent to $\xi$. Therefore,  \eqref{eq:charge-general-0} is only well-defined as the charge associated with $\xi$ at $B$ if it is independent of the choice of the \emph{auxiliary background structure} ${}^{(3)}\!B$. We shall impose this as a consistency condition on the choice of reference term $\bm{B}_X^0$.

\begin{figure}[h!]
	\centering
	\begin{tikzpicture}[ scale=0.8]		
	\draw[black,fill=black] (0,0) circle (.5ex) node [above left] {$B$ };
		\draw[->,thick,>=stealth]  (0,0) to node[at end,right] {$\xi$} ($(0,0)+(2,2)$) ;
	
		\draw[ ] ($(0,0)$)  to[out=45,in=-160] node [at end, below right]    {${}^{(3)}\!B'$} ($(0,0)+(3,1)$);
		\draw[ ] ($(0,0)$)  to[out=-135,in=45]  ($(0,0)-(1,3)$);
		
		\draw ($(0,0)$)  to[out=45,in=-70]  ($(0,0)+(1,3)$);
		\draw ($(0,0)$)  to[out=-135,in=-30] node [at end, left] {${}^{(3)}\!B$} ($(0,0)-(2,2)+(-1,2)$);
	\end{tikzpicture}
	\caption{An example of two hypersurfaces ${}^{(3)}\!B$ and ${}^{(3)}\!B'$ which are tangent at $B$. Both hypersurfaces define the charge \eqref{eq:charge-general-0} associated with $\xi$ at $B$. The consistency condition \eqref{eq:consistency-condition} ensures that the charges are the same for both hypersurfaces.}
	\label{fig:consistency}
\end{figure}
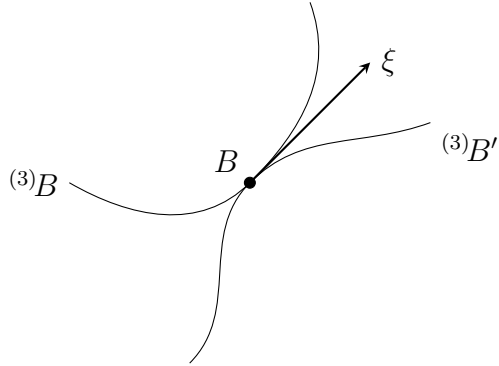

A condition that achieves this is that $\bm{B}_X$ evaluated at $B$ is identical for all ${}^{(3)}\!B$ that are tangent to $\xi$ at $B$. That is, if ${}^{(3)}\!B$ and ${}^{(3)}\!B'$ are any two hypersurfaces that are tangent to each other at $B$, then we require that
\begin{equation}\label{eq:consistency-condition}
\bm{B}_X  \overset{B}{=} \bm{B}_{X}'
\end{equation} 
In this equation, taking the push-forward of $\bm{B}_X$ into the spacetime is understood. See \autoref{fig:consistency} for a situation where this condition should apply.

\subsubsection*{Linearity condition}
\begin{figure}[h!]
	\centering
	\begin{tikzpicture}[ scale=0.6]
	\draw[black,fill=black] (0,0) circle (.5ex) node [right] {$ $};
	
	\draw[ ] ($(0,0)$)  to[out=0,in=180] node [at end,above ]    {${}^{(3)}\!B''$} ($(5,1)$);
	\draw[ ] ($(0,0)$)  to[out=180,in=0] node [at end,above left]    { } ($(-5,-1)$);
	
	\draw[ ] ($(0,0)$)  to[out=45,in=-90] node [at end,above ]    {${}^{(3)}\!B$} ($(2,3)$);
	\draw[ ] ($(0,0)$)  to[out=-135,in=90] node [at end,above left]    { } ($(-2,-3)$);
	
	\draw[ ] ($(0,0)$)  to[out=135,in=-90] node [at end,above ]    {${}^{(3)}\!B'$} ($(-2,3)$);
	\draw[ ] ($(0,0)$)  to[out=-45,in=90] node [at end,above left]    { } ($(2,-3)$);
	
	\draw[->,thick,>=stealth] (0,0) to node [at end, right ]    { $\xi''$} ($(2,0)$);
	\draw[->,thick,>=stealth] (0,0) to node [at end,above  ]    { $\xi'$} ($(-1,1)$);
	\draw[->,thick,>=stealth] (0,0) to node [at end,above ]    { $\xi$} ($(1,1)$);
	\end{tikzpicture}
	\caption{The linearity condition is a constraint on the relation between  the boundary terms $\bm{B}_X$, $\bm{B}_X'$ and $\bm{B}_X''$ on different hypersurfaces that intersect $B$.}
	\label{fig:linearity}
\end{figure}
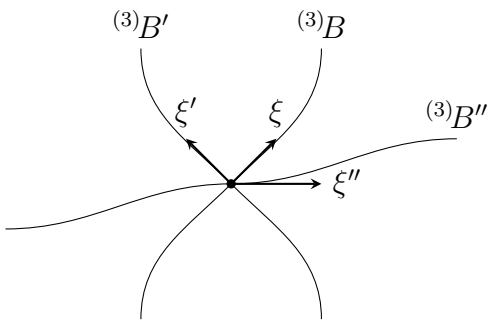

In addition, we shall require that the charge \eqref{eq:charge-general-0} is linear in $\xi$. Thus when $\xi = \xi' + \xi''$, we require that
\begin{equation}
\mathcal{H}_X[\xi] = \mathcal{H}_X [\xi']+\mathcal{H}_X [\xi''].
\end{equation}
Since the Noether charge is linear in $\xi$ \cite{Iyer:1994ys}, it will be sufficient to require that
\begin{equation}\label{eq:linearity-condition}
\xi \cdot \bm{B}_X\overset{B}{=} \xi' \cdot \bm{B}_{X}'+ \xi''\cdot \bm{B}_{X}'', 
\end{equation}
where $\bm{B}_{X}'$ and  $\bm{B}_{X}''$ denote the boundary terms on hypersurfaces tangent to $\xi'$ and $\xi''$ respectively. See \autoref{fig:linearity} for an example where the condition \eqref{eq:linearity-condition} should apply.

A sufficient condition so that both \eqref{eq:consistency-condition} and \eqref{eq:linearity-condition} hold true, is that $\bm{B}_X$ at $B$ is of the form 
\begin{equation}\label{eq:linear-and-consistent}
\bm{B}_X \overset{B}{=} b_X(\phi)\bar{\bm{V}},
\end{equation}
where $\bar{\bm{V}}$ denotes the pull-back of a spacetime three-form $\bm{V}$ (which may depend on $B$) onto ${}^{(3)}\!B$, and $b_X$ is a functional dependent on the boundary data $X$ available on ${}^{(3)}\!B$, but independent of the choice of ${}^{(3)}\!B$.

\subsection{The orthogonality and the zero point conditions}\label{sec:orthogonality-zero-point}
After imposing the condition \eqref{eq:linear-and-consistent}, the freedom left in the choice of $\bm{B}$ is the choice of the three-form $\bm{V}$ and the functional $b_X$. 

We assume that the kernel of the three-form $\bm{V}$ is one-dimensional. We may do this if we absorb multiplicative factors into $b_X$. Then $\bm{V}$ is determined by a direction $\xi_{\perp}$ such that
\begin{equation}\label{eq:orientation}
\xi_{\perp} \cdot \bm{V}  =0.
\end{equation}
 There are two natural choices of $\xi_{\perp}$ at a given closed spacelike two-surface $B$. Namely, the ingoing and outgoing null directions orthogonal to $B$ denoted by $n$ and $l$ respectively. We shall set 
\begin{equation}\label{eq:n-perp}
\xi_{\perp} = n,
\end{equation}
and require that \eqref{eq:orientation} holds true. We refer to this as the \emph{orthogonality condition}.
 
To reduce the freedom in the choice of $b_X$, we shall require that  $\bm{B}_X$ vanishes at $B$ on a reference solution $\phi_0$. I.e.,
\begin{equation}\label{eq:condition-zero}
\bm{B}_X(\phi_0)=0.
\end{equation}
We refer to this as the \emph{zero point condition}.

\subsection{Summary of our proposal}\label{sec:summary-proposal}
Let $\xi$ be a vector field on $M$. Consider a closed spacelike two-surface $B$. Pick a hypersurface ${}^{(3)}\!B$ that contains $B$ and to which $\xi$ is tangent. Denote by $\bm{\Theta}(\phi,\delta \phi)$ a Wald-Zoupas correction term on ${}^{(3)}\!B$. That is, a solution to  \eqref{eq:pull-back-omega}. The charge $\mathcal{H}_X[\xi]$ will then be defined as a solution to \eqref{eq:def-conserved-quantity}. Since $\bm{\Theta}(\phi,\delta \phi)$ is determined up to a total variation $\delta \bm{W}$ in \eqref{eq:form-of-Theta}, our proposal is a  method to specify $\bm{W}$. At this point, we differ from the original prescription by Wald and Zoupas.

We require to choose $\bm{W}=\bm{B}_X$ such that $\bm{\Theta}(\phi,\delta \phi)$ integrates to zero for variations $\delta$ that respect a given choice of boundary conditions $X$ on ${}^{(3)}\!B$. This determines $\bm{\Theta}(\phi,\delta \phi)$ up to a \emph{reference term} $\bm{B}^{0}_X$ which is invariant under variations that respect the boundary conditions $X$.

We reduce the freedom in the choice of reference term by the consistency condition \eqref{eq:consistency-condition}. This condition requires that if we had picked a different ${}^{(3)}\!B$ that contains $B$ and to which $\xi$ is tangent, the quantity $\bm{B}_X$ at $B$ will be the same. In addition to the consistency condition, we require that the charge is linear in the symmetry generators $\xi$. 

To guarantee consistency and linearity, we impose that the boundary term is of the form $\bm{B}_X = b_X(\phi) \bar{\bm{V}}$  (see \eqref{eq:linearity-condition}). Here $\bm{V}$ is a spacetime three-form with a one-dimensional kernel, $\bar{\bm{V}}$ denotes its pull-back onto ${}^{(3)}\!B$, and $b_X(\phi)$ is a functional that depends on the boundary data $X$ available on ${}^{(3)}\!B$, but so that it is independent of the ${}^{(3)}\!B$ that contain $B$.

It then remains to limit the freedom in choosing the functional $b_X$ and the three-form $\bm{V}$. We specify $\bm{V}$ by requiring that for ingoing lightrays generated by $n$ orthogonal to $B$, we have $n\cdot \bm{V}=0$. The functional $b_X$ is required to be chosen so that the \emph{zero point condition} is satisfied, namely that $\bm{B}_X$ vanishes at every $B$ on a reference solution $\phi_0$.

\subsubsection*{Existence and uniqueness}
In \autoref{eq:quasi-local-GR} we shall construct a reference term for the Einstein-Hilbert action, which satisfies \eqref{eq:linear-and-consistent}, \eqref{eq:orientation} and \eqref{eq:condition-zero}, and hence yields a well-defined quasi-local conserved charge.

We want to emphasize that we have not in detail investigated uniqueness (and in general existence) of our proposal. Thus we do not guarantee that our proposal is successful outside of the domain studied in the remainder of the present paper.

\subsubsection*{Computational remark}
In order to evaluate \eqref{eq:charge-general-0} for a given symmetry generator $\xi$ on $B$, it is not required to construct a hypersurface ${}^{(3)}\!B$ that is tangent to $\xi$. Namely,  property \eqref{eq:linear-and-consistent} guarantees that 
\begin{equation}\label{eq:computational-remark}
\xi \cdot \bm{B}_X \overset{B}{=} b_X(\phi) \  \xi \cdot \bm{V},
\end{equation}
so that it is sufficient to determine $\bm{V}$, and to compute $b_X(\phi)$ on a ${}^{(3)}\!B \supset B$ that is convenient. For the reference term that we construct for the Einstein-Hilbert action in \autoref{sec:ref-term}, it is not even necessary to refer to an auxiliary hypersurface ${}^{(3)}\!B$.

\section{Quasi-local charges in General Relativity}\label{eq:quasi-local-GR}
Here we construct quasi-local conserved charges for the four-dimensional ($n=3+1$) Einstein-Hilbert  action according to the prescription in the previous section. The construction  consists of two parts. First, we write down the general form of the charge associated with canonical boundary conditions. Second, we construct a reference term so that the \emph{consistency}, \emph{linearity}, \emph{orthogonality} and \emph{zero point conditions} are satisfied.

\subsection{The form of the Einstein-Hilbert charges}
Consider the Einstein-Hilbert Lagrangian
\begin{equation}
\bm{L}(g) = \frac{1}{16\pi}R(g)\bm{\epsilon}(g),
\end{equation}
where $R(g)$ denotes the Ricci scalar  and $\bm{\epsilon}(g)$ is the volume form associated with $g$. For this theory, a\footnote{The canonical one form $\bm{\theta}$ is defined up to $\bm{\theta}\mapsto \bm{\theta} + \D \bm{Y}$, where $\bm{Y}(\phi,\delta \phi)$ is a covariant $(n-2)$-form linear in $\delta \phi$.} canonical one-form $\bm{\theta}(g,\delta g)$ obtained through the variation of the Lagrangian in \eqref{eq:variation-lagrangian} is given by \cite{Iyer:1994ys}
\begin{equation}\label{eq:theta}
\theta_{abc} = \epsilon_{dabc}v^d,
\end{equation}
where
\begin{equation}
v^d = \nabla_a \delta g^{ad} - \nabla^d (g^{ab}\delta g_{ab}).
\end{equation}
The corresponding Noether charge two-form is given by \cite{Iyer:1994ys}
\begin{equation}\label{eq:Noether-EH} 
\bm{Q}[\xi] := -\frac{1}{16\pi} \epsilon_{abcd} \nabla^{c}\xi^d\D x^a  \D x^b.
\end{equation}
Consider a timelike hypersurface ${}^{(3)}\!B$. Denote the induced metric on ${}^{(3)}\!B$ by $\sigma_{ab}$ and the extrinsic curvature by $K_{ab}$. Then the pull-back of the canonical one-form \eqref{eq:theta} may be expressed as \cite{Iyer:1995kg}
\begin{equation}\label{eq:theta-var-K}
\begin{split}
\bar{\bm{\theta}}_{abc} =-\frac{1}{16\pi}(K^{de}-\sigma^{de}K)&\delta \sigma_{de}\bm{\epsilon}_{abc}-\delta\bigg(\frac{1}{8\pi}K\bm{\epsilon}_{abc} \bigg)+\frac{1}{16\pi} \D \big(m^c \delta m^d \bm{\epsilon}_{abcd} \big),
\end{split}
\end{equation}
where $K:= \sigma^{ab}K_{ab}$, and $m^a$ is the \emph{outward pointing} unit normal vector to ${}^{(3)}\!B$, and the induced volume form is $\bm{\epsilon}_{abc}:= m^d \bm{\epsilon}_{dabc}$.

Now, we write down the form of the correction term defined in \autoref{def:WZ-X} associated with canonical boundary conditions. For the Einstein-Hilbert action, since the variation of the metric appears in $\bm{\theta}(g,\delta g)$, canonical boundary conditions correspond to fixing the induced metric $\sigma_{ab}$ at ${}^{(3)}\!B$. Comparison of \eqref{eq:theta-var-K} with \eqref{eq:theta-B} then immediately tells us that $\bm{B}_X$ for canonical boundary conditions $X$ is given by
\begin{equation}\label{eq:W-non-null}
\bm{B}  = -\frac{1}{8\pi}(\bm{K}-\bm{K}_0),
\end{equation}
where $\bm{K}_0=\bm{K}_0(\sigma)$ is an arbitrary $3$-form functional of the induced metric $\sigma_{ab}$. Therefore, the canonical charge is given by
\begin{equation}\label{eq:charge-general}
	\mathcal{H}[\xi] = \int_{B} \bm{Q}[\xi] +\frac{1}{8\pi}\xi \cdot ( \bm{K}-\bm{K}_0).
\end{equation}

It remains to construct the \emph{reference term} $\bm{K}_0$ satisfying the conditions \eqref{eq:linear-and-consistent}, \eqref{eq:orientation} and \eqref{eq:condition-zero}. This is a non-trivial task. To see why, notice that, for instance, the choice $\bm{K}_0 = 0$ violates the consistency condition \eqref{eq:consistency-condition}. The reference term by Brown and York \cite{Brown:1992br} violates the zero point condition\footnote{This refers to the statement that the Brown-York quasi-local energy does not in general vanish in the Minkowksi spacetime.} \eqref{eq:condition-zero}.

The reference term that we shall construct in \autoref{sec:ref-term} is dependent on a specific formulation of the geometry of $B$ and ${}^{(3)}\!B$, and an expression of $\bm{K}$ therein. We shall define this formulation now.

\subsection{The trace of the extrinsic curvature}\label{sec:trace-K-formulation}
In this section, we provide an expression of $\bm{K}$ in terms of the geometry of a foliation of ${}^{(3)}\!B$ by closed spacelike two surfaces. This expression will be necessary in the next section where we construct a reference term  $\bm{K}_0$ that satisfies the conditions stated in \autoref{sec:conserved-charges}.  We follow the formalism in  \cite{Gourgoulhon:2005ch,Gourgoulhon:2006uc,Booth:2012xm,Booth:2005ss}, which we also refer to for technical details.

\subsubsection*{Evolution vector}
We begin by defining an evolution vector of ${}^{(3)}\!B$.

Let $\Sigma_v$ be a null foliation of the spacetime, labeled by the parameter $v$. Denote by $B_v$ the level surfaces of ${}^{(3)}\!B$ at a constant value of $v$.  Then the \emph{evolution vector} $h$ of  ${}^{(3)}\!B$  is uniquely defined (see \cite{Gourgoulhon:2005ch,Gourgoulhon:2006uc}) by the conditions that (i) $h$ is tangent to ${}^{(3)}\!B$, (ii) $h$ is orthogonal to each $B_v$ and (iii) $\mathcal{L}_h v = 1$. We denote half of the norm of $h$ by
\begin{equation}\label{eq:C}
C:=\frac12 h_a h^a.
\end{equation}

\begin{figure}[h!]
\begin{center}
		\begin{tikzpicture}[ scale=1]
	\draw (0,0) to[out=180,in=0] (-2,-0.5);
	\draw (0,0) to[out=0,in=-140] node [at end, right]    {${}^{(3)}\!B$} (4,-1);

	\coordinate (Bv) at (0,0);
	\draw[->,>=stealth] (Bv) -- ($(Bv)+1.5*(1,1)$) node[at end,above right] {$l$};
	\draw[->,>=stealth] (Bv) -- ($(Bv)+(-1,1)$)node[at end,above left] {$n$};
	\draw[->,>=stealth] (Bv) -- ($(Bv)+3*(1,0)$) node[at end, right] {$h$};
	\draw[dashed,->,,>=stealth] ($(Bv)+1.5*(1,1)$) -- ($(Bv)+1.5*(1,1)+1.5*(1,-1)$) node[midway,above right] {$-Cn$};
	\draw[dashed,->,,>=stealth] ($(Bv)+1.5*(1,1)$) -- ($(Bv)+1.5*(1,1)-1.5*(1,-1)$) node[midway,above right] {$+Cn$};
	\draw[black,fill=black] (Bv) circle (.4ex) node[below] {$\overset{ }{B}_v$};
	\draw[->,>=stealth] (Bv) -- ($(Bv)+1.5*(1,1)-1.5*(1,-1)$) node[at end,above ] {$\tau$};

	\end{tikzpicture}
	\caption{A pictorial representation of the vectors $h$, $\tau$, $l$ and $n$. The norm of these vectors is determined by the foliation $B_v$ of the hypersurface ${}^{(3)}\!B$.}
	\label{fig:vectors}
\end{center}
\end{figure}
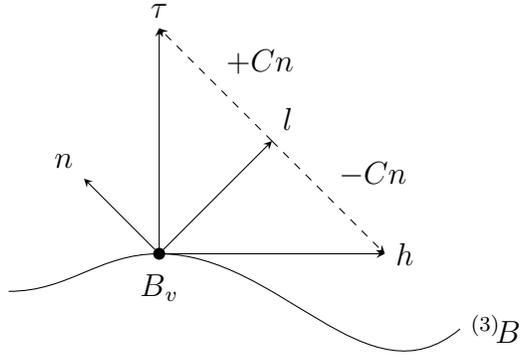

The evolution vector $h$ may be used to define the normalisation of the in- and outgoing null normals orthogonal to $B_v$ denoted by $n$ and $l$ respectively. We normalise them such that $l^a n_a=-1$ and such that
\begin{equation}\label{eq:h-lcn}
h^a = l^a - Cn^a.
\end{equation}
It is then natural to define a vector $\tau$ normal to ${}^{(3)}\!B$ by
\begin{equation}\label{eq:tau}
\tau^a := l^a + C n^a.
\end{equation}
See \autoref{fig:vectors} for a pictorial representation of the vectors defined above.

\subsubsection*{Expansion, surface gravity and the twist form}
The \emph{expansion} of the induced volume element on $B_v$ along the evolution vector is defined as
\begin{equation}\label{eq:expansion-h}
\theta^{(h)}:=\frac12 q^{cd} q^a_cq^b_d \mathcal{L}_h g_{ab}.
\end{equation} 
Here $q_{ab}$ denotes the induced metric on $B_v$. The expansions $\theta^{(\tau)}$, $\theta^{(l)}$ and $\theta^{(n)}$ are  defined similarly. A useful identity is 
\begin{equation}\label{eq:theta-identity}
\theta^{(\tau)} = \theta^{(l)} +C\theta^{(n)}.
\end{equation}
Next, we define a connection on the normal bundle of $B$, referred to as the \emph{twist one-form}, by
\begin{equation}\label{eq:twist}
\omega_a := -n_b \nabla_a l^b.
\end{equation}
The \emph{surface gravity} is defined by
\begin{equation}\label{eq:surface-gravity}
\kappa  := l^a \omega_a.
\end{equation}

\subsubsection*{The trace of the extrinsic curvature}
We are now in a position to express $\bm{K}$ in terms of the quantities defined above. Towards this end, we consider the \emph{trace of the extrinsic curvature with respect to $\tau$}, defined by
\begin{equation}\label{eq:K-tau}
K^{(\tau)}:= \sigma^{ab}\sigma^c_a \sigma^d_b\nabla_c \tau_d.
\end{equation}
Here $\sigma_{ab}$ denotes the induced metric on ${}^{(3)}\!B$. It is related to the induced metric $q_{ab}$ on $B$ by
\begin{equation}
\sigma^{ab}  = q^{ab} + \frac{1}{2C}h^a h^b.
\end{equation}
 Using \eqref{eq:C}, \eqref{eq:h-lcn} and \eqref{eq:surface-gravity}  we may then write
\begin{equation}
K^{(\tau)} =\kappa  +\theta^{(\tau)}-\frac{1}{2C}\mathcal{L}_h C.
\end{equation}
Furthermore, we define on ${}^{(3)}\!B$ the volume form\footnote{The notation $\bar{\bm{V}}$ indicates that we shall later view this volume form as the pull-back of a spacetime three-form.}
\begin{equation}\label{eq:volume-form-ANC}
\bar{\bm{V}}:= \D v \wedge \bm{\epsilon}(q),
\end{equation}
where $\bm{\epsilon}(q)$ denotes the canonical volume form on $B$ associated with the metric $q$. Since on a spacelike  ${}^{(3)}\!B$, we have $\bm{\epsilon}(\sigma) = (2C)^{1/2}\bar{\bm{V}}$, and  \eqref{eq:K-tau} is related to the usual trace of the extrinsic curvature\footnote{The usual trace of the extrinsic curvature is given by \eqref{eq:K-tau} where $\tau$ is replaced by the outward unit normal  vector $m$.} $K$ by $K^{(\tau)} =(2C)^{1/2} K$, it follows that
\begin{equation}\label{eq:K-final}
\bm{K} = -\bigg(\kappa  +\theta^{(\tau)}-\frac{1}{2C}\mathcal{L}_h C \bigg)\bar{\bm{V}}.
\end{equation}
Here, the minus sign arises because the boundary term in \eqref{eq:canonical-action} has the opposite sign for spacelike hypersurfaces. Notice that $\kappa$ and $\theta^{(\tau)}$ in \eqref{eq:K-final} are defined at surfaces of arbitrary signature. The quantity  $C^{-1}\mathcal{L}_h  C$ is not defined at points where the boundary becomes null $C=0$. This issue will be taken care of momentarily. 

\subsection{The reference term}\label{sec:ref-term}
In this section, we construct a reference term $\bm{K}_0$ in \eqref{eq:charge-general} so that the {consistency}, {linearity}, {orthogonality} and {zero point conditions} from \autoref{sec:conserved-charges} are satisfied. For the moment, we assume that ${}^{(3)}\!B$ is everywhere non-null. At the end, we observe that the resulting $\bm{B}_X$ is also well-defined at points of ${}^{(3)}\!B$ which are null.

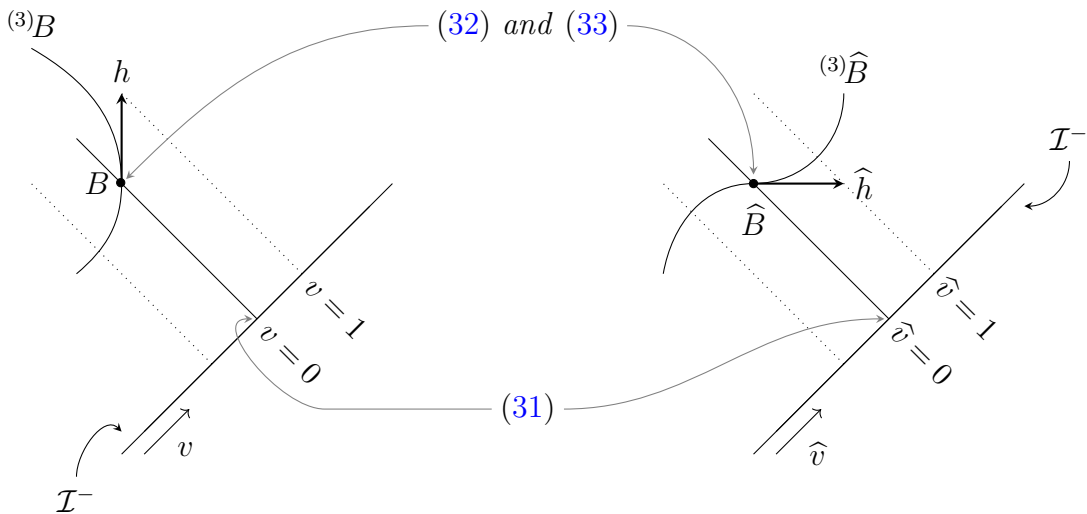
\begin{figure}[h!]
	\centering
	\begin{tikzpicture}[ scale=0.6]


	\coordinate (i-) at (0,0) {};
	\coordinate (i0) at (6,6) {};
	\node (shift) at (1,1) {};
	\draw[] (i-) node [at start,below  ]{$ $} -- node [at end,  right]{$ $} (i0);
	\draw ($ (i-) + 0.5*(shift) $) -- ($ (i0) - 0.5*(shift) $);

	\draw ($(i-)+4*(-1,1)+3*(1,0)$)  to[out=40,in=-90] ($(i-)+(0,6)$);
	\draw  ($(i-)+(0,6)$) to[out=90,in=-30] node [at end, above]    {${}^{(3)}\!B$} ($(i-)+(0,6)+(-2,3)$);

	\draw[->,>=stealth,thick] ($(i-)+(0,6)$) to[out=90,in=-90] node[at end,above] {$h$} ($(i-)+(0,8)$);

	\draw ($(i-)+3*(shift)$) -- node[near start, right ]    {$ $} node [at start, right,rotate =-45]{$v=0$} ($(i-)+3*(shift)+4*(-1,1)$);
	\draw[dotted] ($(i-)+2*(shift)$) -- ($(i-)+2*(shift)+4*(-1,1)$);
	
	\draw[dotted] ($(i-)+4*(shift)$) --node [at start, right,rotate =-45]{$v=1$} ($(i-)+4*(shift)+4*(-1,1)$);

	\draw[black,fill=black] (-0.02,6.02) circle (.5ex) node [left] {$B$};	
	
	\draw[->] ($(i-)+(0.5,0)$) -- node[midway, below right,rotate = 0] {$v$}($(i-)+(0.5,0)+(shift)$);
	
	\coordinate (dot0) at ($(i-)+3*(0,1)$);
	

	\draw[->,>=stealth] ($(i-) + (-1,-0.5)$) to[out=90,in=135] node[at start,below]{$\mathcal{I}^-$} ($(i-) + (0,0.5)$);
	
	
	\coordinate (i-2) at (14,0) {};
	\coordinate (i02) at (20,6) {};
	\node (shift) at (1,1) {};
	\draw[] (i-2) node [at start,below  ]{$ $} -- node [at end,  right]{$ $} (i02);
	\draw ($ (i-2) + 0.5*(shift) $) -- ($ (i02) - 0.5*(shift) $);
	
	\draw ($(i-2)+4*(-1,1)+2*(1,0)$)  to[out=80,in=180] ($(i-2)+(0,6)$);
	\draw  ($(i-2)+(0,6)$)  to[out=0,in=-90] node [at end, above]    {${}^{(3)}\!\hat{B}$} ($(i-2)+(0,6)+2*(shift)$);
	
	\draw[->,>=stealth,thick] ($(i-2)+(0,6)$) to[out=0,in=-180] node[at end,right] {$\hat{h}$} ($(i-2)+(2,6)$);

	\draw ($(i-2)+3*(shift)$) -- node[near start, right ]    {$ $} node [at start, right,rotate =-45]{$\hat{v}=0$} ($(i-2)+3*(shift)+4*(-1,1)$);
	\draw[dotted] ($(i-2)+2*(shift)$) -- ($(i-2)+2*(shift)+4*(-1,1)$);
	
	\draw[dotted] ($(i-2)+4*(shift)$) --node [at start, right,rotate =-45]{$\hat{v}=1$} ($(i-2)+4*(shift)+4*(-1,1)$);

	\draw[black,fill=black] ($(i-2)+(0,6)$) circle (.5ex) node [below] {$\overset{}{\hat{B}}$};	
	
	\draw[->] ($(i-2)+(0.5,0)$) -- node[midway, below right,rotate = 0] {$\hat{v}$}($(i-2)+(0.5,0)+(shift)$);
	
	\coordinate (dot0) at ($(i-2)+3*(0,1)$);
	
	
	\draw[->,>=stealth] ($(i02) + (1,0.5)$) to[out=-90,in=0] node[at start,above ]{$\mathcal{I}^-$} ($(i02) + (0,-0.5)$);
	
	
	\draw[->,>=stealth,color=gray] ($(i-)+3*(shift)+5.2*(1,0)-(0,2)$) -- (4.5,1)  to[in=-180,out=180] ($(i-)+3*(shift)-(0.1,0)$);
	\draw[->,>=stealth,color=gray] ($(i-2)+3*(-1,1)-1.2*(1,0)-(0,2)$) to[in=-180,out=0] ($(i-2)+3*(shift)-(0.1,0)$);
	
	\node (shear)  at (9,1) {\eqref{eq:identify-shear}};
	
	

	\draw[<-,>=stealth,color=gray] ($(i-)+(0.1,6.1)$) to[out=45,in=180] ($(i-)+(0,6)+4*(shift)+(2.8,-0.5)$);
	\draw[<-,>=stealth,color=gray] ($(i-2)+(0,6.2)$) to[out=90,in=0] ($(i-2)+(0,6)+(-2.8,3.5)$);
	
	\node (shear)  at (9,9.5) {\eqref{eq:condition1} \textit{and} \eqref{eq:condition2} };

	\end{tikzpicture}
	\caption{Foliations by Newman-Unti (or Bondi) coordinates $v$ and $\hat{v}$ of the original spacetime $M$ (left) and the reference spacetime $\hat{M}$ (right). The coordinates $v$ and $\hat{v}$ are chosen such that the asymptotic shears at the level surfaces of $v = \hat{v}=0$ are identical (see \eqref{eq:identify-shear}). The surface $B$ is embedded isometrically into the level surface $\hat{v}=0$. Then, a hypersurface ${}^{(3)}\!\hat{B}$ is constructed by the conditions  (\eqref{eq:condition1} and \eqref{eq:condition2}). This identification depends on the evolution vectors $h$ and $\hat{h}$ defined by the foliations of the spacetime.}
	\label{fig:reference}
\end{figure}

We  shall construct $\bm{K}_0$ as the trace of the extrinsic curvature density of ${}^{(3)}\!B$ embedded in a reference spacetime $\hat{M}$. The embedding is completely determined by the intrinsic geometry of ${}^{(3)}\!B$, in agreement with our choice of canonical boundary conditions. The reader may find it helpful to consult \autoref{fig:reference} for a pictorial representation of the construction.\\
\quad \\
\textbf{Step 1} (Spacetime foliation). The first step is to define a (null) foliation of the spacetime. The purpose of this is that a foliation defines the evolution vector in the previous section, on which our formulation of the geometry is dependent. Since we will need to compare the geometry between hypersurfaces in the original and the reference spacetime, we shall need a sensible way to speak about ``the same'' foliation in the original as in the reference spacetime. One place where ``the same'' can be given a meaning is null infinity, since there the geometry of the original and the reference spacetime is the same. Therefore, null infinity will be the place where we shall now set up our spacetime foliations.

We introduce in a neighbourhood of past null infinity $\mathcal{I}^-$ a Newman-Unti\footnote{See \autoref{sec:NU}  for a review of Newman-Unti coordinates.} \cite{Newman:1962cia} (or if the reader prefers a Bondi \cite{Bondi:1962px}) coordinate system $(v,r,x^A)$. The coordinate $v$ labels a foliation of $M$ by null hypersurfaces $\Sigma_{v}$. The coordinate $r$ parametrises\footnote{In Newman-Unti gauge, $r$ is an affine parameter of the geodesics generated by $n$. In Bondi gauge, $r$ is the \emph{areal} or \emph{luminosity distance}.} the null geodesic generators of $\Sigma_v$. The angular coordinates $x^A$ label the null generators of $\Sigma_v$. The asymptotic metric in these coordinates is given by
\begin{equation}
\D s^2 = -  \D v ^2 + 2 \D v \D r +(r^2 \gamma_{AB}+r C_{AB}) \D x^A \D x^B + (...).
\end{equation}
(See \autoref{sec:quasi-local-BMS} for more precise asymptotic conditions.) There are infinitely many of such Newman-Unti (or Bondi) coordinates $(v,r,x^A)$, labelled by the \emph{asymptotic shear} $C_{AB}$. We  consider the Newman-Unti coordinates such that $B$ is entirely contained in the null hypersurface\footnote{It is not necessary that the Bondi coordinates cover the surface $B$. Namely, the null hypersurface $\Sigma_v$ is defined independent of the coordinate $r$.} $\Sigma_0$. 

We then define $\hat{\Sigma}_0$ in the reference spacetime $\hat{M}$ as the level surface of a Newman-Unti coordinate $\hat{v}$, such that the corresponding asymptotic shear $\hat{C}_{AB}$ at $\hat{v} =0$ is identical to $C_{AB}$ at $v =0$, i.e.,
\begin{equation}\label{eq:identify-shear}
\hat{C}_{AB}|_{\hat{v}=0} = C_{AB}|_{v=0}.
\end{equation}
 \quad \\
\textbf{Step 2} (Isometric embedding). We embed $B$ isometrically by a map $i:B\hookrightarrow \hat{\Sigma}_0$. The image of $B$ is denoted by $\hat{B}:=i[B]$.\\
\quad \\
\textbf{Step 3} (Constructing ${}^{(3)}\! \hat{B}$). Denote by ${}^{(3)}\!\hat{B}$ at this point any hypersurface that contains $\hat{B}$ and consider its foliation by the null hypersurfaces $\hat{\Sigma}_{\hat{v}}$. Let $\hat{h}$ be the corresponding evolution vector defined in the previous section and denote by $\hat{n}$ the corresponding ingoing null normal. As defining conditions of  ${}^{(3)}\!\hat{B}$, we then require that
\begin{equation}\label{eq:condition1}
C\theta^{(n)}=i^* \big(\hat{C} \theta^{(\hat{n})} \big),
\end{equation}
and
\begin{equation}\label{eq:condition2}
C^{-1}\mathcal{L}_h C = i^* \big( \hat{C}^{-1}\mathcal{L}_{\hat{h}} \hat{C} \big).
\end{equation}
Here $i^*$ denotes the pull-back of the map $i$ defined in step 2, and $\hat{C}$ is defined by \eqref{eq:C} for the evolution vector $\hat{h}$. (Notice that $C\theta^{(n)}$ and $C^{-1}\mathcal{L}_h  C$ are the quantities in \eqref{eq:K-final} which depend on the choice of ${}^{(3)}\!B$. Therefore, \eqref{eq:condition1} and \eqref{eq:condition2} will ensure that the resulting reference term satisfies the consistency condition.) \\
\quad \\
\textbf{Step 4} (Embedding of ${}^{(3)}\!B$ into ${}^{(3)}\!\hat{B}$). Extend the map $i:{}^{(3)}\!B \hookrightarrow {}^{(3)}\!\hat{B}$ around $B$ such that $i^*(\hat{v}) = v$. In other words, the extension is defined by identifying the Newman-Unti coordiates $v$ and $\hat{v}$. This extension is not unique -- it may be twisted off B -- but since the resulting reference term in \eqref{eq:ref-term-evaluated} will not depend on this freedom, we do not fix it. \\
\quad \\
\textbf{Step 5} (The reference term). Finally, define at $B$
\begin{equation}\label{eq:K0-final}
\bm{K}_0 :\overset{B}{=} i^*\bm{\hat{K}},
\end{equation}
where $\hat{\bm{K}}$ denotes the extrinsic curvature density \eqref{eq:K-final} of ${}^{(3)}\!\hat{B}$. Comparison of \eqref{eq:K-final} and \eqref{eq:K0-final} with \eqref{eq:condition1}, \eqref{eq:condition2} and  \eqref{eq:theta-identity}, yields that
\begin{equation}\label{eq:ref-term-evaluated}
\bm{K}-\bm{K}_0 \overset{B}{=}-  \big(\kappa -\hat{\kappa} +\theta^{(l)} -\hat{\theta^{(l)}} \big) \bar{\bm{V}}.
\end{equation} 
Here, in our notation we denote by $\hat{\circ}$ the reference value of the quantity $\circ$.

 Notice that, in contrast to \eqref{eq:K-final}, \eqref{eq:ref-term-evaluated} is also well-defined at points where ${}^{(3)}\!B$ becomes null ($C=0$). This is because the divergent piece at null surfaces in \eqref{eq:K-final} was identified in the reference spacetime by \eqref{eq:condition2}.

\subsection{Consistency, linearity, orthogonality and the zero point }
Here we show that \eqref{eq:ref-term-evaluated} is of the form \eqref{eq:linear-and-consistent}.

First, notice that the volume form \eqref{eq:volume-form-ANC} is at $B$ the pull-back of the spacetime three-form
\begin{equation}
\bm{V} := -n_a \D x^a \wedge \bm{\epsilon}(q).
\end{equation}
(The values of $\bm{V}$ outside of $B$ are irrelevant for our purposes.) From this, it follows that 
\begin{equation}
n \cdot \bm{V}=0,
\end{equation}
so that $\bm{V}$ satisfies the orthogonality condition \eqref{eq:orientation} with respect to the ingoing light direction $n$.  Second, note that $\theta^{(l)}$ and $\kappa$ depend only on $B$. This means that the term between brackets in \eqref{eq:ref-term-evaluated} is independent of ${}^{(3)}\!B\supset B$. Therefore,  \eqref{eq:linear-and-consistent} is satisfied. This proves consistency and linearity. Lastly, the zero point condition \eqref{eq:condition-zero} is trivially satisfied.

\subsection{Existence and uniqueness}
Here we comment on the existence and uniqueness of the reference term as constructed in \eqref{eq:ref-term-evaluated}.

As an example, consider Minkowski space as the reference spacetime and suppose that $B$ is contained in a slice $\Sigma_0$ for which the asymptotic shear vanishes: $C_{AB}|_{v=0}=0$. Then the (degenerate) metric on $\hat{\Sigma}_0$ is given by
\begin{equation}
\D s^2 = 0 \D r^2 + r^2 \D \Omega^2.
\end{equation}
It follows directly from the \emph{uniformization theorem}\footnote{The uniformization theorem states that every metric on $S^2$ is conformal to the round metric.} that the embedding map $i$ exists and is unique up to isometries of the Minkowski spacetime. It then remains to construct ${}^{(3)}\!\hat{B}$.  Towards this end, let ${}^{(3)}\!\hat{B}$ be located at\footnote{Here $(\hat{v},\hat{r},\hat{x}^A)$ denote Newman-Unti coordinates of the Minkowski spacetime with $\hat{C}_{AB}=0$.} $\hat{r} = p(\hat{v},\hat{x}^A)$. Then \eqref{eq:condition1} uniquely determines  $\partial_{\hat{v}} p|_B$ and \eqref{eq:condition2} uniquely determines $\partial^2_{\hat{v}} p|_B$. (The function $p|_B$ is determined by the embedding $i$.) This determines the trace of the extrinsic curvature of the hypersurface ${}^{(3)}\!\hat{B}$ at $B$. Therefore, the reference term exists and is unique.

We leave a more general study of existence and uniqueness for future work.

\subsection{The canonical Einstein-Hilbert charges}\label{eq:evaluate-EH}
Finally, we  evaluate the canonical quasi-local charges for the Einstein-Hilbert Lagrangian.

A given vector field $\xi$ at $B$ may be decomposed as
\begin{equation}\label{eq:decomposition}
\xi = \alpha l + \beta n + \xi_{||}.
\end{equation}	
Here $n$ and $l$ are the in- and outgoing null normals\footnote{The normalisation of these null normals is defined in \autoref{sec:trace-K-formulation} where the parameter $v$ is now a Newman-Unti coordinate.} to $B$, and $\xi_{||}$ is a vector tangent to $B$. In terms of the decomposition \eqref{eq:decomposition}, the pull-back of the Noether charge two-form \eqref{eq:Noether-EH} onto $B$ becomes\footnote{To derive this expression, we used that
	\begin{equation}
	\bm{\epsilon} = \bm{l} \wedge \bm{n} \wedge  \bm{\epsilon}(q).
	\end{equation}}
\begin{equation}\label{eq:Noether-EH-evaluated}
\overline{\bm{Q}[\xi]}|_{B}=\frac{1}{16\pi}\big( \alpha \kappa + \mathcal{L}_l\alpha -\mathcal{L}_n \beta+ 2\xi^a_{||} \bar{\omega}_a \big) \bm{\epsilon}(q),
\end{equation}
where $\kappa$ is the surface gravity defined in \eqref{eq:surface-gravity} and $\bar{\omega}_a$ denotes the pull-back of the twist form \eqref{eq:twist} onto $B$. Substitution of \eqref{eq:Noether-EH-evaluated} and \eqref{eq:ref-term-evaluated} into \eqref{eq:charge-general} then yields that
\begin{equation}\label{eq:charge-evaluated-with-K0}
	\begin{split}
	\mathcal{H}[\xi] &= \frac{1}{16\pi}\int_{B}\bigg[ \alpha \kappa    + \mathcal{L}_l\alpha -\mathcal{L}_n \beta+2 \xi^a_{||}\bar{\omega}_a -2 \alpha \big( \theta^{(l)}-\hat{\theta^{(l)}} +\kappa- \hat{\kappa}\big) \bigg] \bm{\epsilon}(q).
	\end{split}
\end{equation}
This concludes the construction of canonical quasi-local charges for the Einstein-Hilbert action.

\section{Quasi-local BMS charges}\label{sec:quasi-local-BMS}
In the previous sections, we provided a consistent method to define a conserved charge associated with a symmetry generator $\xi$ at a closed spacelike two-surface $B$. Our next task is to consider a specific symmetry generator $\xi$ to evaluate the charge. Here our choice\footnote{One other natural choice would be $\xi = l$. However, we do not consider this choice here, because its associated charge does not vanish in the Minkowski spacetime. } will be that $\xi$ is a BMS vector field \cite{Bondi:1962px,Sachs:1962wk,Sachs:1962zza}. 

Usually, BMS symmetries are only considered in the asymptotic region. Namely, they are defined as diffeomorphisms that act non-trivially at null infinity, but which preserve the asymptotically flat boundary conditions. Their action in the bulk is generally considered arbitrary and therefore irrelevant.

However, there do exist ways to extend BMS symmetries into the bulk of a spacetime. For example, when a gauge such as Bondi gauge or Newman-Unti gauge has been fixed, the extension of BMS generators into the bulk is unique by the requirement that they preserve the given gauge conditions. 

This does, however, not take away the problem that the gauge fixing method is essentially arbitrary. Different gauge fixing methods lead to different BMS generators. And unfortunately, our charge \eqref{eq:charge-evaluated-with-K0} is -- in the bulk of the spacetime -- not independent\footnote{This problem also occurs at null infinity. Usually it is assumed that the representatives satisfy the \emph{Geroch-Winicour condition}, which guarantees uniqueness of the charge. See e.g. \cite{Wald:1999wa}.} of this gauge choice. (This may be verified by comparison of the BMS charge in the Newman-Unti and Bondi gauges.)

In order to define BMS charges, one must make a choice of gauge. Our choice will be Newman-Unti gauge. Our motivation for this choice is that the BMS generators in Newman-Unti gauge are connected to the \emph{gravitational memory effect} in the bulk of a spacetime. This connection is explained in a separate paper \cite{bart2019}, which generalises the observation of Strominger and Zhiboedov \cite{Strominger:2014pwa} that BMS symmetries at null infinity are connected to gravitational memory. 

The organisaton of this section is as follows. After a review of BMS symmetries in Newman-Unti gauge in \autoref{sec:NU}, we evaluate the associated \emph{quasi-local BMS charges} in \autoref{sec:EH-charges}. Then we show in \autoref{sec:BMS-Minkowski} that the BMS charges vanish in the Minkowski spacetime, and in \autoref{sec:asympt} that they yield the correct asymptotic behaviour at null infinity. In \autoref{sec:BMS-spherical} we compute the charges in the Vaidya and Reissner-N\"ordstrom spacetimes, in order to argue in \autoref{sec:BMS-energy} that the zero mode BMS charge  is a promising definition of \emph{quasi-local energy}. 

\subsection{BMS generators in Newman-Unti gauge}\label{sec:NU}

Newman-Unti coordinates \cite{Newman:1962cia} are based on a null foliation of the spacetime parametrised by the first coordinate $v$. The second coordinate $r$ is an affine parameter for the null geodesic generators $n_a = -\partial_a v$ in the hypersurfaces $\Sigma_v$ of constant $v$. The remaining angular coordinates $x^A$ are defined such that $n^a$ generates light rays at constant angles.

The metric in these coordinates takes the form
\begin{equation}\label{eq:NU-form}
\D s^2 = W \D v^2 + 2 \D r \D v + g_{AB}  (\D x^A - V^A \D v)(\D x^B - V^B \D v).
\end{equation}
Part of the freedom left in the choice of $(v,{r},x^A)$ is then used to impose the following\footnote{One way to obtain these expressions is to consider the fall-off conditions in Bondi gauge \cite{Barnich:2011mi} and to use the relation between the Bondi and Newman-Unti gauges given by Equation (4.5) of \cite{Barnich:2011ty}.} fall-off conditions. Namely, 
\begin{equation}\label{eq:asymptotic-metric}
g_{AB} = \big( {r}^2-4\beta_0 \big) \gamma_{AB} + {r} C_{AB}  +D_{AB}+ O(r^{-1}),
\end{equation}
where $\gamma_{AB}$ is the round metric, $\partial_v D_{AB} = 0$  and $C_{AB}$ is traceless with respect to $\gamma_{AB}$,
\begin{equation}\label{eq:traceless}
\gamma^{AB}C_{AB}=0,
\end{equation}
and
\begin{equation}\label{eq:beta0}
\beta_0  :=  -\frac{1}{32}C_{AB}C^{AB}.
\end{equation}
(In the previous section $C_{AB}$ was referred to as the asymptotic shear of the geodesic null congruence defined by $n^a$.) Furthermore,
\begin{equation}
V^A = \overset{\infty}{V}r^{-2} +O(r^{-3}),
\end{equation}
where 
\begin{equation}
\overset{\infty}{V} := \frac12 \overset{\circ}{D}_B C^{AB}.
\end{equation}
Here $\overset{\circ}{D}_A$ is the covariant derivative with respect to the round metric $\gamma_{AB}$. And
\begin{equation}
W = -1 +\frac{2m_B+4 \partial_v\beta_0}{r} +O(r^{-2}),
\end{equation}
where $m_B$ denotes the \emph{Bondi mass aspect}. The inverse metric is given by
\begin{align}
g^{ab}=\begin{pmatrix}
0& 1&0 \\
1 & -W &  V^A \\
0& V^B &  g^{AB}
\end{pmatrix}.
\end{align}

BMS symmetries are diffeomorphisms that preserve the asymptotic fall-off conditions of an asymptotically flat spacetime. In Newman-Unti gauge, they are generated by\footnote{In $3+1$ dimensions.} vector fields $\xi$ of the form \cite{Newman:1962cia,Barnich:2011ty}
	\begin{equation}\label{eq:BMS}
	\begin{cases}
	\xi^{v} = f\\
	\xi^r = J-r\partial_v f + \frac12 \overset{\circ}{\Delta}f\\
	\xi^A = Y^A +I^A\\
	\end{cases}
	\end{equation}
	where
	\begin{equation} 
	\begin{split}
	f &:= T(x^A) + \frac12 v \overset{\circ}{D}_A Y^A,\\
	I^A &:= -\partial_B f \int^{r}_\infty  g^{AB} \D r',\\
	J &:=-\partial_Af\int_{\infty}^r V^A \D r'.
	\end{split}
	\end{equation}
	Here $T(x^A)$ is an arbitrary function of the angular coordinates, referred to as a \emph{supertranslation}, and $Y^A$ is a conformal Killing vector of $\gamma_{AB}$. The operator  $\overset{\circ}{\Delta}:=\overset{\circ}{D}_A\overset{\circ}{D}{}^A$ denotes the spherical Laplacian.

\subsubsection*{Domain of applicability}
The BMS generators \eqref{eq:BMS} are only defined at the points in the spacetime where Newman-Unti coordinates are defined. When the curvature of the spacetime becomes too strong, the lightrays in $\Sigma_v$ generated by $n^a$ start to intersect\footnote{The light rays intersect when $\theta^{(n)} =  \nabla_a n^a=0$.}, at which point the coordinates become ill-defined.

\begin{figure}[h!]
	\centering
	\begin{tikzpicture}[ scale=0.7]
	\draw (0,0)  circle (1.5);
	\node (I)    at (0,0)   { };
	
	\filldraw[lightgray] (-3.5,0)  circle (0.5);	
	\draw[dashed] (-6,0) -- (0,0);
	\draw[dashed] (-6,0.9) to[in = 160,out=-5] (0,-0.2);
	\draw[dashed] (-6,-0.9) to[in = 200,out=5] (0,0.2);
	\end{tikzpicture}
	\caption{A planet in the presence of a black hole. The trajectory of ingoing light rays generated by $n^a$ is deformed by the planet. However, if the energy momentum distribution of the planet is sufficiently weak, the ingoing light rays intersect inside of the trapping region.}
	\label{fig:bh-planet}
\end{figure}
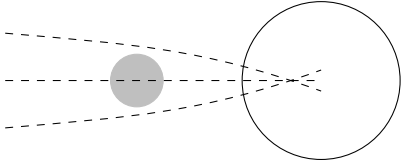

However, Newman-Unti coordinates do cover many interesting situations. We illustrate this with an example. Consider a planet in the vicinity of a black hole. When the energy density of the planet is sufficiently small or the planet is sufficiently close to the black hole, the light rays generated by $n^a$ intersect behind the horizon. This means that the above BMS generators are defined at black hole horizons with sufficiently weakly gravitating matter in the exterior. See \autoref{fig:bh-planet}.

\subsection{Einstein-Hilbert charges}\label{sec:EH-charges}
Here we evaluate \eqref{eq:charge-evaluated-with-K0} for the case that $\xi$ is a BMS generator \eqref{eq:BMS}.

Consider a two-surface $B$ located at \emph{constant} values of $v$ and $r$ in Newman-Unti coordinates. The in- and outgoing null normals $n$ and $l$ respectively are then given by
\begin{equation}
n^a = (0,-1,0,0)\quad \text{and} \quad l^a = \bigg(1,-\frac{W}{2}, V^A\bigg) .
\end{equation} 
The BMS vector field \eqref{eq:BMS} in the decomposition \eqref{eq:decomposition} is given by
\begin{align}
\alpha &= f,\\
\beta &= -\xi^r -\frac{\alpha W}{2},\\
\xi_{||}^A &= \xi^A - \alpha V^A.
\end{align}
This yields
\begin{align}
\mathcal{L}_n \beta &= -\mathcal{L}_l f - \alpha \kappa,
\end{align}
where we used that $\kappa = -\frac12\partial_r W $. The charge may now be seen to evaluate to
\begin{equation}\label{eq:BMS-charge-evaluated-with-K0}
	\begin{split}
	\mathcal{H}[\xi] &= \frac{1}{8\pi}\int_{B}\bigg[ -f \big( \theta^{(l)}-\hat{{\theta}^{({l})}}+V^A {\omega}_A - \hat{\kappa} \big)+ \mathcal{L}_l f + \xi^A{\omega}_A  \bigg]\bm{\epsilon}(q).
	\end{split}
\end{equation}
This is the main result of this section.

The quantity $\hat{\kappa}$ in \eqref{eq:BMS-charge-evaluated-with-K0} may be interpreted as a reference term for $V^A \omega_A$. To see this, one may verify that in the Minkowski spacetime in Newman-Unti coordinates it holds true that
\begin{equation}\label{eq:k-equals-VOmega}
\kappa = V^A \omega_A \quad (\text{in the Minkowski spacetime}).
\end{equation}
The quantity $\xi^A$ is a geodesic deviation between the light rays generated by $n^a$ and light rays that are BMS deformations thereof. This observation is elaborated in \cite{bart2019}.

In the following, we show  (i) that the BMS charges \eqref{eq:BMS-charge-evaluated-with-K0} vanish in the Minkowski spacetime, and (ii) that they coincide asymptotically at null infinity with the BMS charges known in the literature.

\subsection{Vanishing charges in the Minkowski spacetime}\label{sec:BMS-Minkowski}
Here we show that the quasi-local BMS charges vanish in the Minkowski spacetime. Since the Minkowski spacetime is also our reference spacetime that satisfies the zero point condition \eqref{eq:condition-zero}, it follows from \eqref{eq:charge-general-0} that the conserved charge vanishes when the Noether charge vanishes. We now show that the Noether charge associated with BMS generators vanishes in the Minkowski spacetime.

Consider the Minkowksi spacetime given by
\begin{equation}\label{eq:minkowski}
\D s^2 = - \D v^2 + 2 \D v \D r + r^2 \gamma_{AB} \D x^A \D x^B.
\end{equation}
One may verify that on an arbitrary closed two-surface $B$ the pull-back of $\bm{Q}[\zeta]$  (defined in \eqref{eq:Noether-EH}) onto $B$ is a total derivative on $B$ when $\zeta$ satisfies
\begin{align}
\partial_r \zeta^v &= 0,\\
\partial_v \zeta^v - \partial_r \zeta^r &= \overset{\circ}{D}_A (\ \cdot\ )^A,
\end{align}
and
\begin{equation}
\zeta^A = \frac{1}{r} \gamma^{AB}\partial_B \lambda,
\end{equation}
where $\partial_v \lambda = \partial_r \lambda = 0$. Furthermore, in Minkowski space, the Noether charge two-form $\bm{Q}[k]$ also integrates to zero on $B$ when $k$ is an isometry.

Since in Minkowski space in Newman-Unti gauge it holds true that BMS vector fields are a linear combination of the vector fields $\zeta$ (supertranslations) and isometries $k$ (rotations and boosts), the corresponding Noether charge vanishes\footnote{The same reasoning holds true for BMS generators in Bondi gauge. See \cite{Bondi:1962px,Barnich:2011mi} for the expression of the BMS generators in Bondi gauge.}.

\subsection{Asymptotic behaviour}\label{sec:asympt}
Here we show that the asymptotic limit of the quasi-local BMS charges \eqref{eq:BMS-charge-evaluated-with-K0} agrees with the BMS charges constructed by Wald and Zoupas \cite{Wald:1999wa} at null infinity. In our notation, we denote by $\mathcal{C}_v$ a cut at null infinity at a constant value of $v$, i.e., the limit of $B$ as $r\rightarrow \infty$.

Notice first that $\omega^A = \frac12 \partial_r V^A$. Since
\begin{equation}
V^A\omega_A = O(r^{-3}),
\end{equation}
this term, and  because of \eqref{eq:k-equals-VOmega} also its reference value $\hat{\kappa}$ in \eqref{eq:charge-evaluated-with-K0} do not contribute to the charge as $r\rightarrow \infty$. Second, one may verify that the asymptotic value of the term containing $I^A \omega_A + \mathcal{L}_l f$ vanishes.

Next, we compute $\theta^{(l)}$ and its reference value $\hat{\theta^{(l)}}$. The asymptotic expansion of $\theta^{(l)}$ is given by
\begin{equation}\label{eq:theta-l-infty}
\theta^{(l)} = \frac{1}{r}+ \frac{1}{r^2}\big( \overset{\circ}{D}_A \overset{\infty}{V}{}^A- 2 m_B \big)+O(r^{-3}),
\end{equation}
The reference value $\hat{\theta^{(l)}} $ is given by \eqref{eq:theta-l-infty} where the quantities $C_{AB}$ and $m_B$ are replaced by their reference values $\hat{C}_{AB}$ and $\hat{m}_B$. From our construction of the reference term in \autoref{sec:ref-term}, it follows that (at the given value of $v$)
\begin{align}
\hat{C}_{AB}  &=C_{AB},\\
\hat{m}_B &= 0,
\end{align}
so that the asymptotic value of the charge is given by
\begin{equation}\label{eq:desired-form}
\overset{\infty}{\mathcal{H}}[\xi] = \frac{1}{8\pi}\int_{\mathcal{C}_v} \big(2f m_B  + Y^A N_A \big) \bm{\epsilon}(\gamma).
\end{equation}
Here $N_A$ denotes the \emph{angular momentum aspect} which we define as the subleading coefficient in the asymptotic expansion of $\omega_A$, given by
\begin{equation}
\omega_A =- \frac{1}{2r} \overset{\circ}{D}{}^BC_{AB} + \frac{1}{r^2}N_A +O(r^{-3}).
\end{equation}
(The leading order term of $\omega_A$ does not contribute to the charge. This follows from the facts that $\overset{\circ}{D}{}^BC_{AB}$ is a total derivative on $\mathcal{C}_v$, $Y^A$ is a conformal Killing vector and the trace of $C_{AB}$ vanishes.)

The asymptotic charge \eqref{eq:desired-form} is the desired form. See for comparison\footnote{The definition of $N_A$ in \cite{Barnich:2011mi}, which we denote by $N_A'$, is related to ours by $N_A = N_A' -  \partial_A \beta_0$, where $\beta_0$ was defined in \eqref{eq:beta0}. } e.g. Equation (3.2) of \cite{Barnich:2011mi}.

\subsection{The charge in spherically symmetric spacetimes}\label{sec:BMS-spherical}
In this section, we compute the BMS charges \eqref{eq:BMS-charge-evaluated-with-K0} on spherically symmetric spacetimes given by \eqref{eq:NU-form}, where
\begin{equation}\label{eq:spherical}
\begin{split}
W &= W(v,r),\\
V^A &= 0, \\
g_{AB} &= r^2\gamma_{AB}.
\end{split}
\end{equation}
Here $\gamma_{AB}$ is the round metric.

The only contribution to the charge comes from the null expansion $\theta^{(l)}$ and its reference value. For a surface $B$ at constant $(v,r)$, they are given by
\begin{align}
\theta^{(l)} &= -\frac{W}{r},\\
\hat{\theta^{(l)}} &=  \frac{1}{r}.
\end{align}
The resulting charges are
\begin{equation}\label{eq:charge-spherical}
\mathcal{H}[\xi] =\frac{1}{8\pi} \int_{B}  \frac{f (1+W(v,r))}{r}\bm{\epsilon}(q).
\end{equation}
Only the zero mode (time translation) $f=1$  contributes to the charge, for which it is equal to the \emph{Misner-Sharp} energy \cite{Misner:1964je,Cahill1970,Szabados:2009eka} 
\begin{equation}
E_{\text{MS}}(v,r) := \frac{r}{2}(1+W(v,r)).
\end{equation}

\subsubsection*{Vaidya metric}
The Vaidya metric is given by the Schwarzschild metric where the mass parameter is made time dependent. That is, \eqref{eq:spherical} where
\begin{equation}
W(v,r)=-\bigg( 1 - \frac{2m(v)}{r}\bigg).
\end{equation}
It describes the formation of a black hole by a spherically symmetric shell of null dust. For $B$ at arbitrary radii, we find that the gravitational part of the canonical charge is given by
\begin{equation}
\mathcal{H}[f=1] = m(v).
\end{equation}

\subsubsection*{Reissner-N\"ordstrom black hole}
The Reissner-N\"ordstrom metric is given by \eqref{eq:spherical} where
\begin{equation}\label{eq:W-RN}
W(v,r)= - \frac{(r-r_-)(r-r_+)}{r^2},
\end{equation}
where
\begin{equation}
r_{\pm}:=m\pm \sqrt{m^2-Q^2}.
\end{equation}
We find that
\begin{equation}
\mathcal{H}[f=1] = \frac{ (r_++r_-)r-r_+r_-}{2r}.
\end{equation}
In particular, at $r = r_+$, we have
\begin{equation}
\mathcal{H}[f=1]|_{r=r_+} = \frac{r_+}{2}.
\end{equation}
This quantity is equal to the \emph{irreducible mass} of the black hole, given by
\begin{equation}\label{eq:Mirr}
m_{\text{irr}} := \sqrt{\frac{\text{Horizon Area}}{16\pi}}, \quad \text{where \quad Horizon Area $= 4\pi r_+^2$}.
\end{equation}

Notice that the Reissner-N\"ordstrom black hole is a solution to the vacuum Einstein-Maxwell equations, not the Einstein-Hilbert equations. Therefore, the charge also receives a contribution from the gauge field. Although the inclusion of gauge fields in \autoref{sec:conserved-charges} is straight forward, it happens to be the case that the resulting charges are not independent of the electromagnetic gauge. This gauge dependence follows from the fact that the symplectic two form is not gauge invariant \cite{Wald:1999wa}. One may have to substantially fix the gauge  -- or choose appropriate boundary conditions -- in order for our procedure to yield gauge invariant quantities. We have left these investigations for the future.

\subsection{Quasi-local energy}\label{sec:BMS-energy}
Consider the BMS charge given by \eqref{eq:charge-evaluated-with-K0} where $\xi$ is the $Y^A=0$, $f=1$ BMS vector field \eqref{eq:BMS}. Summarizing the observations of the previous section, this charge has the following properties:
\begin{enumerate}
\item It vanishes on the Minkowski spacetime,
\item It asymptotes to the Bondi mass at null infinity,
\item On the round spheres in the metric \eqref{eq:spherical} it is equal to the Misner-Sharp energy. Specifically, at  the outer horizon of a Reissner-N\"ordstrom black hole, it is the \emph{irreducible mass}.
\end{enumerate}
These properties are contained in a list of pragmatic criteria that a reasonable notion of quasi-local energy is expected to satisfy \cite{Szabados:2009eka}. We therefore put forward the possibility that the \emph{gravitational part} of the zero mode BMS charge as constructed above may be a useful definition of quasi-local energy.

\section{Concluding remarks}\label{sec:discussions}
We provided a general construction of quasi-local ``conserved'' charges in General Relativity. The construction may be thought of as a modification of the prescription of Wald and Zoupas for defining conserved quantities at null infinity. Our modification is applicable more generally, and in particular in the bulk of a spacetime. We applied our construction to BMS symmetries in the bulk of  asymptotically flat spacetimes, so as to define quasi-local BMS charges. We then argued that the zero mode BMS charge is a promising definition of quasi-local energy. 

Let us conclude with the following remarks.

(i) Because of computational complexity, we did not consider the Kerr geometry in our examples in \autoref{sec:quasi-local-BMS}. However, the expression of the Kerr metrics in Newman-Unti gauge is known \cite{Fletcher2003}. Therefore, our BMS charges are in principle also defined in the Kerr spacetime. It would be useful to check if the zero mode BMS charge at the outer horizon of a  Kerr black hole is equal to the irreducible mass.

(ii) In stating that the gravitational part of the zero mode BMS charge at the horizon of a Reissner-N\"ordstrom black hole is the irreducible mass, we purposefully ignored the contribution from the gauge field to the canonical charge. We did this, because the contribution from the gauge field is dependent on the electromagnetic gauge. In order to construct a gauge invariant quantity, one could consider different boundary conditions, such as fixing the electric charge instead of the gauge field. Another possibility would be to fix the gauge substantially, perhaps similar to the way we fixed Newman-Unti gauge for BMS generators. We have left this for future investigations.

(iii) Our prescription may be used to define quasi-local conserved charges in spacetimes with different asymptotic conditions. 

(iv) Our prescription in \autoref{sec:conserved-charges} is formally applicable to diffeomorphism covariant theories in general. However, we have not investigated this in any detail.

\section*{Acknowledgements}
I thank  Glenn Barnich, Gerard Bart, Lasha Berezhiani,  Ivan Booth, Geoffrey Compere,  Marc Henneaux, Florian Hopfm\"uller, Dieter L\"ust, Stefan Vandoren and in particular David Osten and Sebastian Zell for useful discussions. I thank Dieter L\"ust, David Osten and in particular Sebastian Zell for feedback on the draft.

\appendix

\end{spacing}

\bibliographystyle{JHEP}

\bibliography{references}

\end{document}